\documentclass[prb,amsmath,amssymb,12pt,showpacs]{revtex4}
\usepackage{latexsym}
\usepackage{graphics}
\usepackage{amssymb}
\usepackage{bbm}
\def\Eloc{{\cal E}}
\def\H{{\mathcal H}}
\def\MC{Monte Carlo}
\def\alfa{\alpha}
\def\bhat{\hat{\beta}}
\def\det{{\rm det\,}}
\def\ds{ }
\def\half{{1\over 2}}
\def\mat#1{{#1}}
\def\v#1#2{{\rm\bf #1}_{#2}}
\def\R{{\rm\bf R}}
\def\N{{\bf N}}
\def\H{{\bf H}}
\def\Nhat{{\bf \hat N}}
\def\Hhat{{\bf \hat H}}

\begin{document}
\setlength{\topmargin}{0.3cm}
\title{ Inter-dimensional Degeneracies in van der Waals Clusters and
  Quantum Monte Carlo Computation of Rovibrational States}
\author{M. P.  Nightingale}
\affiliation{ Department of Physics, University of Rhode Island, Kingston
  RI 02881, USA}
\author{Mervlyn Moodley}  
\affiliation{School of Physics, Howard College, University of KwaZulu-Natal, 
  Durban, 4041, South Africa} 
\date{\today}
\begin{abstract}
Quantum \MC\ estimates of the spectrum of rotationally invariant
states of noble gas clusters suggest inter-dimensional degeneracy
in $N-1$ and $N+1$ spacial dimensions.  We derive
this property by mapping the Schr\"odinger eigenvalue problem onto
an eigenvalue equation in which $D$ appears as a continuous variable.
We discuss implications for quantum \MC\ and dimensional scaling
methods.
\end{abstract}
\maketitle

\section{\label{sec.intro}Introduction}

One of the advantages of \MC\ methods is that they scale well with the
number of degrees of freedom of a physical system.  In this paper we
consider van der Waals clusters consisting of $N$ bosonic
Lennard-Jones atoms in $D$ spatial dimensions.  We treat the atoms as
``elementary'' particles without internal degrees of freedom, so that
in total we deal with clusters with $ND$ quantum mechanical degrees of freedom.  We are
mainly interested in the energy spectra of these clusters.

Quantum \MC\ computations can be made much more efficient by the use of
optimized trial wave functions, as is well known.  With currently available methods,
as a matter of fact, the problem of
computing rovibrational spectra with \MC\ methods is virtually
intractable without good trial functions.  One of the questions of
interest is the relative importance of the quality of these trial
wave functions for $n$-body correlations with $n$ in excess of the
commonly used correlations with $n=2$ and $n=3$.  In this context,
the idea of varying the spatial dimensionality of the system quite naturally
suggests itself, because particles can be more compact in higher
dimensions, which suggests that correlations involving a higher number
of particles might become more important as the spatial
dimensionality increases.  While we have not found clear numerical
evidence to support this idea,\cite{Merv} our computations did produce
an interesting by-product, which forms the topic of this paper.

Our computations showed that the energy spectra of $N$ particles in
$N-1$ and $N+1$ spatial dimensions are numerically indistinguishable
for states invariant under rotation and translation.\cite{Merv}
Indeed, in this paper we show that for these $S$-states and for $D\ge
N-1$, the $N$ particle time-independent Schr\"odinger equation can be
transformed into an eigenvalue equation involving a differential
operator with ${1\over 2}(N-1)N$ independent variables and an
effective potential in which the spatial dimension $D$ appears as a
continuously varying parameter.  This effective potential turns out
to depend quadratically on $D$ and is symmetric about $D=N$. This
implies the aforementioned inter-dimensional degeneracy, in addition to
a relationship between the wave functions for $D=N-1$ and $D=N+1$
dimensions.  We note here that, as is also manifest in the frustration
of the classical system for $D<N-1$ ---frustration for example in the
sense that not all inter-atomic distances can be equal--- the spectrum
for values of $D<N-1$ cannot be obtained by analytic continuation of
the spectrum for $D\ge N-1$; we shall return to this in the discussion
at the end of this paper.

Inter-dimensional degeneracy was also derived recently by Gu {\it et
  al,}\cite{Gu1} by a group-theoretical method.
To the best of our
knowledge, inter-dimensional degeneracies of $S$-states of
Lennard-Jones clusters have not been observed before, with the
exception of the two-body cluster in one and three dimensions
which follows from the standard separation of variable solution of the
Schr\"odinger equation.

Even though an extension of our study to dimensions higher than the
physical three dimensions is primarily of academic interest, the
effect of spatial dimension on quantum systems has been studied since
the early days of quantum physics. In fact, Fock \cite{Fock} as early
as 1935 showed that there exists a relationship between the
the hydrogen-like wave functions and four-dimensional hyper-spherical
harmonics.\cite{Avery} The
hyper-spherical coordinate method was used in the late seventies to
discover inter-dimensional degeneracies in electron systems. For the
one-electron system a transformation was found that
reveals inter-dimensional degeneracy between a system in
$D$ dimension and angular momentum $l$ with the same system in $D \pm
2$ dimensions and angular momentum $l \mp 1$.\cite{Herrick1}
Many-electron systems have also been shown to exhibit inter-dimensional
degeneracies.\cite{HerrickSt} Further references to other
inter-dimensional studies can be found in Ref.~\onlinecite{Gu1}.

The structure of this paper is as follows. In Sec.~\ref{sec.opt} we
briefly summarize the quantum Monte Carlo technique for excited
states developed in Refs.~\onlinecite{NV1} and \onlinecite{NV2} to
obtain optimized trial wave functions for van der Waals clusters. In
this method, trial wave functions are developed that can be improved
systematically. These trial wave functions are linear combinations of
elementary basis functions with non-linear variational
parameters. The elementary basis functions consist of a prefactor and
exponential polynomial that is formulated in terms of all possible
$N$-body correlations. In Sec.~\ref{sec.numres} we present Monte Carlo
energy estimates obtained for selected few-body van der Waals clusters
in a limited number of dimensions ranging from $D=1$ to $D=6$. In this
study we consider van der Waal clusters composed of atoms of Kr, Ar,
Ne and the hypothetical $\frac{1}{2}$-Ne, which has half the
(dimensionless) mass of Ne. Kr can be considered as a
semi-classical case while that of $\frac{1}{2}$-Ne is more
quantum mechanical in nature.  Sec.~\ref{sec.exact} is devoted to the exact
derivation of dimensional degeneracy, with some of the results
postponed to the Appendix.  In the final Section~\ref{sec.disc} we
discuss the relevance of our results, in particular for dimensional scaling
methods.

\section{\label{sec.mc}\MC\ approach}
\subsection{\label{sec.opt}Optimization of ground and excited state wave 
functions}
We consider clusters in $D$ dimensions consisting of $N$ atoms with
positions specified by the $D\times N$ matrix of Cartesian coordinates
$\R=(\v r 1 \v r 2 \dots \v r N)$, with
\begin{equation}
\v r i=\left(\begin{array}{c}
x_{1i}\\
\vdots\\
x_{Di}
\end{array}\right).
\label{eq.r}
\end{equation}
We shall use the following definitions
\begin{subequations}
\begin{eqnarray}
\v r {ij}&=& \v r j - \v r i\label{eq.diff}\\
r_{ij} &=&|\v r {ij}|
\end{eqnarray}
\end{subequations}
for difference vectors and their lengths.

For a system of $N$ bosonic van der Waals atoms with atomic mass $\mu$
and interacting via a pair potential, the dimensionless Hamiltonian is
\begin{equation} \label{Ham1}
H=-\frac{1}{2m}\sum_{i=1}^{N} \nabla_i^2 + \sum_{(i,j)} V(r_{ij}),
\end{equation}
with
\begin{equation}
\nabla_i^2=\sum_{\alpha=1}^D {\partial^2\over\partial x_{\alpha i}^2}
\end{equation}
and where $V$ is the dimensionless
Lennard-Jones potential
\begin{equation}
V(r) = {1 \over r^{12}} - {2 \over r^6}.
\end{equation}
The inverse dimensionless mass is given by $m^{-1} = \hbar^2
/2^{\frac{1}{3}} \mu \sigma^2 \epsilon$, which is proportional to the
square of the de Boer parameter,\cite{Boer} where $-\epsilon$ is the
minimum of the Lennard-Jones potential and $2^{1\over 6}\sigma$ the
corresponding inter-particle distance.

A preliminary step in our optimization procedure is to generate a
sample of configurations ${\bf R}_\sigma$, with $\sigma=1, \dots ,s,$
which are sampled from a relative probability density function
$\psi_g({\bf R}_\sigma)^2$. The guiding function $\psi_g$ used for the
computations reported in this paper is defined in terms of a trial
function $\tilde \psi$, which approximates the ground state.
In simple cases, we used $\psi_g^2=\tilde \psi^{2/\rho}$, where
the parameter $\rho$ is chosen in the range $2\alt\rho\alt
3$.  Where necessary, we used a more sophisticated guiding
function\cite{unpublished} so as to generate a sample
with substantial overlap with all the excited states under
consideration.

The trial wave functions are linear combinations of elementary basis
functions $\beta_i$, each of which implicitly depends on non-linear
variational parameters,  and we use different procedures to optimize the
linear and non-linear parameters. For reasons explained in detail below, we
define the re-weighted functions $\bhat_i(R)= \psi_g(R)^{-1}\beta_i(R)$
and $\bhat_i'(R)= \psi_g(R)^{-1} H \beta_i(R)$.  For a complete set of
elementary basis functions $\beta_i$ the Schr\"odinger equation can be
written in the form
\begin{equation} \label{HamBH}
\bhat_i'({\bf R_\sigma})=\sum_{j=1}^n \bhat_j({\bf R_\sigma}) \Eloc_{ji}.
\end{equation}
In practical applications, the set of functions $\beta_i$ is, of
course, far from complete, but the $n\times n$ matrix $\Eloc$ may
still be determined by solving Eq.~(\ref{HamBH}) for $\Eloc$ in
a least-squares sense given the re-weighting just introduced.  Note that
Eq.~(\ref{HamBH}) is exactly satisfied if the functions $\beta_i$ span
an invariant subspace of the Hamiltonian $H$, even if they do not form a
{\it complete} set; this provides an important zero-variance principle for
the corresponding part energy spectrum.

It is convenient to rewrite Eq.~(\ref{HamBH}) in matrix form
\begin{equation} \label{mat.eq}
{\bf B'}= {\bf B\Eloc},
\end{equation}
where $B_{\sigma i}=\bhat_i({R_\sigma})$ and $B'_{\sigma
  i}=\bhat'_i({R_\sigma})$. Multiplying Eq.~(\ref{mat.eq}) from the
left by the transpose of $\bf B$, one obtains by inversion
\begin{equation} \label{ENH.eq}
{\bf \Eloc}=({\bf B}^T{\bf B})^{-1}({\bf B}^T{\bf B'})\equiv 
\Nhat^{-1}\Hhat,
\end{equation}
with ${\hat N_{ij}}=\sum_ \sigma \bhat_i(R_ \sigma) \bhat_j(R_
\sigma)$ and ${\hat H_{ij}}=\sum_ \sigma \bhat_i(R_ \sigma)
\bhat_j'(R_ \sigma)$. As can be easily verified, Eq.~(\ref{ENH.eq}) is
indeed the least-squares solution of Eq.~(\ref{HamBH}).  Note that for
an infinite sample the hermiticity of the Hamiltonian guarantees that
$\Hhat$ is a symmetric matrix, but this is not the case for a finite
\MC\ sample.  If $\Hhat$ is symmetrized in Eq.~(\ref{ENH.eq}), the
resulting $\Eloc$ no longer satisfies the least-squares property
nor the aforementioned zero-variance principle.

The optimal linear combinations of the basis functions $\beta_i$ are
computed by constructing the spectral decomposition of $\Eloc$:
\begin{equation}
\Eloc_{ij}=\sum_{k=1}^n d_i^k {\tilde E_k}{\hat d_j^k}
\end{equation}
where $\hat d_j^k$ and $d_i^k$ are the components of the left and
right eigenvectors of $\Eloc$ with eigenvalues $\tilde E_k$. This
yields the trial functions
\begin{equation}
\tilde\psi^k=\sum_{i=1}^n d_i^k \beta_i.
\label{eq.psi-tilde}
\end{equation}

Before we continue this review of our optimization procedure, some
comments should be made.  First of all, the matrix ${\bf \hat N}$
frequently is ill-conditioned.  This problem can be dealt with by
using a singular value decomposition of the matrix ${\bf B}$ to obtain
a numerically regularized inverse of $\Nhat$.\cite{NV1,NV2} Secondly,
we mention that Eq.~(\ref{eq.psi-tilde}) can also be derived from the
condition that the quantum mechanical expectation value of the
Hamiltonian in the states $\tilde\psi^k$ is stationary with respect to
variation of the coefficients $d^k_i$.  This condition yields a
generalized eigenvalue equation involving matrices $\N$ and $\H$, the \MC\
estimators of which are the matrices $\Nhat$ and $\Hhat$ introduced
previously.  The re-weighting defined before Eq.~(\ref{HamBH}) was
introduced so that these estimators are unbiased.

As mentioned, the basis function $\beta_i$ depend implicitly on
non-linear variational parameters.  These are optimized iteratively
and it should be kept in mind that for each choice of the {\it
  non-linear} parameters, new optimized {\it linear} parameters have
to be computed.  The full optimization of {\it all} parameters
therefore entails a linear optimization nested in a non-linear one.
The linear optimization is a standard linear algebra problem;
the optimization of the non-linear parameters is performed by
minimizing the variance of the local energy of the wave function:
\begin{equation}
\chi^2={\ds\sum_{\sigma=1}^s [{\hat \psi}^k{'}({\bf R}_\sigma)-
{\tilde E}_k {\hat \psi}^k ({\bf R}_\sigma)]^2 \over
\ds\sum_{\sigma=1}^s {\hat \psi}^k({\bf R}_\sigma)^2},
\end{equation}
where ${\hat \psi}^k=\psi_g^{-1}{\tilde \psi}^k$ and ${\hat
\psi}^k{'}=\psi_g^{-1}H{\tilde \psi}^k$.

The trial wave functions produced by this method yield estimates of the energy
levels that are upper bounds to the exact energies, if statistical errors
are negligible.  To reduce the resulting systematic errors, the so-called
{\it variational errors,} we employ these optimized wave functions as basis
functions in a correlation function Monte Carlo calculation.\cite{CB1,BCL1,BGL}
This reduces the variational bias in the eigenvalue estimates, but it usually
increases the statistical errors in the estimates.  In a formal sense, the
reduction of variational errors obtained in correlation function \MC\
is accomplished by introducing a new and improved basis by means of the substitution
\begin{equation}
\beta_i(R) \to \exp(-tH) \beta_i(R)\equiv\beta_i(R,t).
\end{equation}

For increasing projection time $t$ the spectral weight of more highly excited
states in the new basis is reduced, and with it the variational error.
In the limit $t\to\infty$ all states of the new basis collapse onto the
ground state, which implies that as $t$ increases, the overlap matrix of the
$t$-dependent basis states becomes more nearly singular,  which increases the
statistical errors.  In principle, the errors increase exponentially; in practice,
the method as we currently use it, breaks down once the \MC\ estimate of the
overlap matrix develops negative eigenvalues.

We use elementary basis functions of the following general form\cite{MMN}
\begin{equation} \label{basisgen.eq}
\beta_i(R)=s_i(R)\exp (\sum_j a_j s_j(R) +
\sum_{\sigma < \tau}A(r_{\sigma \tau})),
\end{equation}
where the term involving $A$ imposes short- and long-range asymptotics;
the $s_i$ and $s_j$ are bosonically symmetrized monomials.  The exact
structure of these basis functions is of no concern in this paper. A
detailed description of the above mentioned method and the structure
of the basis functions can be found in Refs.~\onlinecite{NV1,NV2}.

\subsection{\label{sec.numres} Numerical results in various dimensions}

In this section we present numerical results that show that the energy
spectrum as a function of dimensionality for $D\ge N-1$ is symmetric
about $D=N$.  We discuss results for  Kr, Ar,
Ne and the hypothetical $\frac{1}{2}$-Ne, which
are defined respectively by the following inverse masses:
$1.9128\times 10^{-4}$ (Kr)
$6.9635\times 10^{-4}$ (Ar)
$7.0920\times 10^{-3}$ (Ne)
$1.4184\times 10^{-2}$ ($\frac{1}{2}$-Ne).

\subsubsection{The three-body case}
Table~\ref{GSdim} shows the ground state energies $E_1$ for Kr$_3$,
Ar$_3$ and $\frac{1}{2}$-Ne$_3$ in dimensions ranging from $D=1$ to
$D=6$.\begin{table}[!hpt]
\begin{center}
\begin{tabular}
{|c|
c|r@{$\times$}l|
c|r@{$\times$}l|
c|r@{$\times$}l|}
\hline
&
\multicolumn{3}{c|}{Kr$_3$} &
\multicolumn{3}{c|}{Ar$_3$} &
\multicolumn{3}{c|}{$\frac{1}{2}$-Ne$_3$}\\
\hline
\multicolumn{1}{|c|}{$D$}&
\multicolumn{1}{c|}{$E_1$}&\multicolumn{2}{c|}{$\Delta E_1$}&
\multicolumn{1}{c|}{$E_1$}&\multicolumn{2}{c|}{$\Delta E_1$}&
\multicolumn{1}{c|}{$E_1$}&\multicolumn{2}{c|}{$\Delta E_1$}\\
\hline
1&-1.872~548~547~6&-9&$10^{-1} $&-1.734~808~71 &-8&$10^{-1}$& -0.895~584 
&-4&$10^{-1}$\\
2&-2.760~461~351~5& 2&$10^{-10}$&-2.552~953~22 &-1&$10^{-9}$& -1.302~484 
&-7&$10^{-7}$\\
3&-2.760~555~278~7& 6&$10^{-10}$&-2.553~289~43 & 1&$10^{-8}$& -1.308~442 & 
9&$10^{-6}$\\
4&-2.760~461~351~3&-5&$10^{-11}$&-2.552~953~22 &-1&$10^{-9}$& -1.302~483 
&-2&$10^{-6}$\\
5&-2.760~179~569~8&-1&$10^{-9} $&-2.551~944~61 &-2&$10^{-8}$& -1.284~627 
&-1&$10^{-5}$\\
6&-2.759~709~937~6& 5&$10^{-10}$&-2.550~263~64 & 7&$10^{-9}$& -1.254~901 & 
5&$10^{-6}$\\
\hline
\end{tabular}
\end{center}
\caption{Ground state energies $E_1$ (with errors in the least
significant digit) and deviations from quadratic fits $\Delta E_1$
for Kr$_3$, Ar$_3$ and $\frac{1}{2}$-Ne$_3$ in dimensions $D=1$
through $D=6$.}
\label{GSdim}
\end{table}
We fitted the computed values for $D\ge 2$ to a parabola with its
minimum at $D=2$.  The difference between the computed and fitted
results $\Delta E_1$ is also shown in the Table~\ref{GSdim}.  As is
the case with the classical minimum of the energy, which equals
$-2.03$ for $D=1$ and $-3$ for $D\ge 2$, the quantum mechanical ground
state energy at $D=1$ is nowhere near the curve.

\subsubsection{The four-body case}
Table~\ref{GS_Ar4_dim} shows the ground state energies $E_1$ for a
four-body cluster, Ar$_4$, in various dimensions. From these results
it is evident that an inter-dimensional degeneracy exists in $D=3$ and
$D=5$ dimensions.  Again, $\Delta E_1$ represents the difference of
the computed energies and the results obtained from a parabolic fit with
its minimum at $D=N=4$, this time excluding $D=1$ and $D=2$. The quantum
mechanical estimates can be compared with
the classical minima, viz., $-3.07$ for $D=1$, $-5.07$ for $D=2$,
and $-6$ for $D\ge 3$.

\begin{table}[!hpt]
\begin{center}
\begin{tabular}{|c|c|r@{$\times$}l|}
\hline
\multicolumn{1}{|c|}{$D$} &
\multicolumn{1}{c|}{$E_1$} &
\multicolumn{2}{c|}{$\Delta E_1$}\\
\hline
1       & -2.625~622~56 & -2& $10^{-0}$ \\
2       & -4.329~517~95 & -8& $10^{-1}$ \\
3       & -5.118~146~05 & -2& $10^{-9}$ \\
4       & -5.118~653~84 &  3& $10^{-9}$ \\
5       & -5.118~146~05 & -2& $10^{-9}$ \\
6       & -5.116~622~70 &  1& $10^{-9}$ \\
\hline
\end{tabular}
\end{center}
\caption{Ground state energies (with errors in the least significant digit)
and deviations from quadratic fits $\Delta E_1$ for Ar$_4$ in dimensions
$D=1$ through $D=6$.}
\label{GS_Ar4_dim}
\end{table}

\subsubsection{Excited states}
Thus far we have only numerically verified that inter-dimensional degeneracies
exist for ground state energies. Table~\ref{ExS_Ar3_dim} and
\ref{ExS_Ar4_dim} strongly suggest that the same holds for excited
states. The first table shows the four lowest excited state
energies obtained for Ar$_3$ in $D=2,3$ and $4$ dimensions.  Once
again, as observed for the ground state energies, these degeneracies
exist in this three-body cluster for the $D=2$ and $D=4$ case. The
$D=3$ case is included in this table to indicate that the energies
obtained here are different and lower than the other two cases. The
second table shows the four lowest excited state energies obtained for
Ar$_4$ in $D=3$ and $5$ dimensions;  we denote energy levels
by $E_1<E_2<\cdots$.

\begin{table}[!hpt]
\begin{center}
\begin{tabular}{|l|l|l|l|}
\hline
$k$  &\hskip 5pt $D=2$ & \hskip 5pt $D=3$ &\hskip 5pt $D=4$ \\
\hline
2   &-2.249~860~2   &-2.250~185~5  & -2.249~860   \\
3   &-2.126~038~8   &-2.126~361    & -2.126~039    \\
4   &-1.996~153     &-1.996~43     & -1.996~153    \\
5   &-1.946~3       &-1.946~7      & -1.946~3    \\
\hline
\end{tabular}
\end{center}
\caption{Comparison of the excited state energies $E_k$ (with errors
in the least significant digit) of Ar$_3$ in $D=2,3$ and $4$ dimensions.}
\label{ExS_Ar3_dim}
\end{table}

\begin{table}[!hpt]
\begin{center}
\begin{tabular}{|l|l|l|l|}
\hline
$k$  &\hskip 5pt $D=3$ & \hskip 5pt $D=5$  \\
\hline
2   &  -4.800~897~73  & -4.800~897~75  \\
3   &  -4.725~156~7   &-4.725~156~6    \\
4   &   -4.630~025    & -4.630~025     \\
5   &    -4.586~389   & -4.586~384     \\
\hline
\end{tabular}
\end{center}
\caption{Comparison of the excited state energies $E_k$ (with errors
in the least significant digit) of Ar$_4$ in $D=3$ and $5$ dimensions.}
\label{ExS_Ar4_dim}
\end{table}

The results in Table\ref{ExS_Ne5_dim} illustrate the loss of accuracy
that occurs for five particle clusters. The differences between the
estimates of the energies of corresponding levels for four and six
dimensions are due to the failure to converge of the correlation function
\MC.  This reflects the fact that our trial wave functions
can in principle be systematically improved only for cluster sizes
$N\le 4$, because they contain fully adjustable $n$-body correlations
with $n\le 4$ only.

\begin{table}[!hpt]
\begin{center}
\begin{tabular}{|l|l|l|l|}
\hline
$k$  &\hskip 5pt $D=4$ & \hskip 5pt $D=6$  \\
\hline
1   &  -5.821~21      & -5.821~21     \\
2   &  -5.346~6	      & -5.337~2      \\
3   &  -5.26          & -5.18         \\
4   &  -5.06          & -4.99         \\
5   &  -4.95          & -4.91         \\
\hline
\end{tabular}
\end{center}
\caption{Comparison of ground and excited state energies $E_k$ 
(with uncontrolled errors) of Ne$_5$ in $D=4$ and $6$ dimensions.}
\label{ExS_Ne5_dim}
\end{table}

\section{\label{sec.exact}Exact results}
\subsection{Clusters in arbitrary number of dimension}

The Schr\"odinger equation for an $N$ particle cluster in $D$ spatial
dimensions is a differential equation in $ND$ variables.  For a $S$
state the wave function is invariant under rotations and translations.
Therefore, one can write the wave function as a function of fewer than
$ND$ variables.  To accomplish this we proceed as follows.

Consider the $N-1$ difference vectors $\v r {21},\v r {31},\dots,\v r
{N1}$ as defined by Eq.~(\ref{eq.diff}).  Note that these vectors cannot
be linearly independent unless $D\ge N-1$, in which case they define
a parallelepiped $P$. Precisely $\half(N-1)N$ independent
variables are required to define $P$ up to a congruence
transformation.  One possible choice of such variables consists
of: {\it(1)} the angles $\theta_{ij}$ between the the vectors $\v r {i1}$ and
$\v r {j1}$ or their cosines
\begin{equation}
g_{ij}=\frac{\v r {1i}\cdot\v r {1j}}{r_{1i} r_{1j}}
\end{equation}
with $1<i<j\le N$; and {\it(2)} the lengths of the vectors $\v r
{i1}$ with $1<i\le N$.  Alternatively, as independent variables one
may choose the lengths of all distinct inter-particle distances
$r_{ij}=r_{ji}$ with $i\ne j$.  These are the variables we shall use
in this paper with the assumption, required for linear independence, 
that $D\ge N-1$.

\subsection{Generalized Schr\"odinger equation}

We consider a $D$-dimensional Schr\"odinger equation of the form
\begin{equation}
\left(-\sum_{i=1}^N {1\over 2m_i} \nabla_i^2+V\right)\psi=E\psi
\end{equation}
with a Hamiltonian slightly more general than the one defined in
Eq.~(\ref{Ham1}) with a potential that is rotationally and
translationally invariant, but not necessarily a sum of two-body
contributions.  Furthermore, the mass of each particle may be
different.

We restrict ourselves to $S$ states and to cases in which $D\ge N-1$
so that, as discussed in the previous subsection, the wave functions can
be considered to be a function of {\it independent} inter-particle
distances $r_{ij}$ with $1<i<j\le N$.

By straightforward application of the differential operator identity
\begin{equation}
{\partial \over \partial x_{\alfa i}}=
\sum_{j\ne i} {\partial r_{ij}\over  \partial x_{\alfa i}}{\partial \over 
\partial r_{ij}}
\end{equation}
one obtains
\begin{equation}
\nabla_i^2=\sum_{j\ne i} a_{i;j}{\partial \over \partial r_{ij}}+
\sum_{j,k\ne i} g_{i;jk}{\partial^2\over\partial r_{ij} \partial r_{ik}}
\label{eqn.nablai}
\end{equation}
where
\begin{equation}
a_{i;j}=\sum_{\alfa=1}^D {\partial^2 r_{ij}\over \partial x_{\alfa 
i}^2}={D-1\over r_{ij}},
\end{equation}
and
\begin{equation}
g_{i;jk}=\sum_{\alfa=1}^D
{\partial r_{ij}\over \partial x_{\alfa i}} {\partial r_{ik}\over \partial 
x_{\alfa i}}=
{\v r {ij}\cdot\v r {ik}\over r_{ij}r_{ik}}.
\label{eq.gijk}
\end{equation}

With the inter-particle distances as independent variables, the
Schr\"odinger equation assumes a form that involves: {\it(1)} a linear
differential operator that explicitly depends on the spatial
dimensionality $D$; and {\it(2)} a second-order differential and a
potential energy operator that are independent of $D$, as is clear
from Eqs.~(\ref{eqn.nablai}) and (\ref{eq.gijk}).

Next, we transform the Schr\"odinger equation into an equation in which
the second-order operator is unchanged, the linear operator is absent,
and in which the potential is modified by an additional term.\cite{footnote}
This is accomplished as follows:
\begin{equation}
\H\psi=E\psi \to \H'\phi=E\phi
\end{equation}
with
\begin{equation}
\psi=\chi\phi,
\end{equation}
and
\begin{equation}
\H'={1\over \chi}\H\chi.
\end{equation}
The action of the operator on the right-hand side of an
arbitrary function is to be evaluated from right to left, so that
multiplying by $\chi$ takes precedence over operating by $\H$.

This yields a special case of Eq.~(3.8) of
Ref.~\onlinecite{AveryGoodsonHerschbach}
\begin{equation}
\H'=V-\sum_{i=1} {1\over 2 m_i}(S_i+T_i+U_i)
\end{equation}
with
\begin{equation}
S_i=\sum_{j,k\ne i} g_{i;jk}{\ds\partial^2\over\ds\partial r_{ij} \partial 
r_{ik}},
\end{equation}
\begin{equation}
T_i=\sum_{j\ne i} \left(a_{i;j}+
2\sum_{k\ne i} g_{i;jk}\chi^{-1} {\ds\partial\chi\over\ds \partial 
r_{ik}}\right)
{\ds\partial\over\ds\partial r_{ij}},
\end{equation}
\begin{equation}
U_i=\sum_{j\ne i} a_{i;j}\chi^{-1} {\partial\chi\over\partial r_{ij}}+
\sum_{j,k\ne i} g_{i;jk}\chi^{-1} {\ds\partial^2\chi\over\ds \partial r_{ik} 
\partial r_{ik}}.
\label{eq.U}
\end{equation}

We define square matrices of order $N-1$,
\begin{equation}
\hat{\mat G_i}=(r_{ij} g_{i;jk} r_{ik})_{j,k\ne i}
\end{equation}
for $i=1,\dots,N$.  The matrix $\hat{\mat G_i}$ is the Grammian
associated with the $N-1$ vectors $\v r {ij}$ with
$j=1,\dots,i-1,i+1,\dots,N$.  Its determinant
\begin{equation}
\omega=\det(\hat{\mat G_i})
\end{equation}
is the square of the volume of the parallelepiped defined by the
vectors pointing from particle $i$ to all other particles.  This volume
is equal to $N!$ times the volume of the $(N-1)$-simplex of which the
$N$ particles are the vertices, which explains why $\omega$ does not
depend on $i$, as our notation indicates.

In the Appendix we show that $T_i$ vanishes for the choice
\begin{equation}
\chi=\omega^{(1-D)/4},
\label{eq.chi}
\end{equation}
while
\begin{subequations}
\begin{eqnarray}
U_i&=&
{1\over 8} [(N-1)^2-(N-D)^2]\sum_{j \ne i} {1\over r_{ij}}
{\partial \log \omega\over\partial r_{ij}}\\
&=&{(N-1)^2-(N-D)^2\over 16 \omega^2}\sum_{j,k\ne i}
{\partial \omega\over \partial r_{ij}} g_{i;jk}
{\partial \omega\over \partial r_{ik}}.
\label{eq.Und}
\end{eqnarray}
\end{subequations}

Clearly, $U_i$ depends on the spatial dimensionality via its $D$
dependent amplitude, which is symmetric in $D$ about $D=N$.  Recalling
that this derivation is valid only for values for $D \ge N-1$, we find
that for $S$ states the Schr\"odinger equation has the same energy
eigenvalues in $D=N-1$ and $D=N+1$ ---and for those values of $D$ only---
while the eigenstates are related via
\begin{equation}
%{\psi(D=N-1)\over\chi(D=N-1)}={\psi(D=N+1)\over\chi(D=N+1)}.
\psi(D=N-1)=\sqrt\omega\, \psi(D=N+1).
\end{equation}
Furthermore, using the fact that $g_{i;jk}$, defined in
Eq.~(\ref{eq.gijk}), is an inner product, one can rewrite the sum in
the Eq.~(\ref{eq.Und}) as a sum of squares.  This implies that to
linear order in perturbation theory the energy eigenvalues depend
quadratically on $D$ with a minimum at $D=N$, in agreement with our
numerical estimates presented in Section \ref{sec.numres}.

%\title{Stuff}
%\section{cosines}
%\begin{equation}
%\gamma_{ijk}=\cos\theta_{ijk}
%\end{equation}
%\begin{equation}
%{\partial \gamma_{ijk}\over r_{ik}}=-{r_{ik}\over r_{ij} r_{jk}}
%\end{equation}
%\begin{equation}
%{\partial \gamma_{ijk}\over r_{ij}}=-{r_{ik}\over r_{ij} r_{jk}}\gamma_{jik}=
%{1\over r_{jk}}-{1\over r_{ij}}\gamma_{ijk}
%\end{equation}

\section{\label{sec.disc}Discussion}

We transformed the Schr\"odinger equation for the rotationally and
translationally invariant states of an $N$-particle cluster in $D\ge
N-1$ spatial dimensions into a differential equation in ${1\over
  2}(N-1)N$ independent variables in which the dependence on $D$ is
fully contained in an effective potential energy. Here $D$ can
be interpreted as a continuously varying parameter, as is commonly
done in dimensional scaling studies.\cite{ShiKais} In agreement with
work by Gu {\it et al.,}\cite{Gu1} we observed that there exists an
inter-dimensional degeneracy, of an $N$-body cluster in $D=N-1$ and
$D=N+1$ dimensions. This degeneracy exists for all $S$ states, i.e.,
for both the ground and excited states. Furthermore, the minimum energy was
observed to be at the dimension $D=N$.

We stumbled upon this inter-dimensional degeneracy numerically by \MC\
methods, for which the generalization to arbitrary discrete dimensions
is simple.  In retrospect, knowing that this degeneracy is an exact
property of the Schr\"odinger equation is useful because it provides a
powerful check of the validity of our computer code and of our estimates
of systematic and statistical errors in our numerical results.

Our results have another interesting implication.  Our transformation
of the Schr\"odinger equation to a differential equation in ${1\over
  2}(N-1)N$ independent variables is valid only for $D\ge N-1$.
However, $D$ in the resulting equation can be interpreted as a
continuous variable, and the equation has an analytical continuation
for $D<N-1$ and is symmetric about $D=N$.  As a consequence, the
energy spectrum for $D< N-1$ of the transformed equation is
analytic in $D$ and symmetric about $D=N$.  Whatever might be the
meaning of this spectrum obtained by analytic continuation of the
spectrum for large values of the $D$ spectrum, it cannot have anything to
do with the physical spectrum of $N$ particle clusters for $D< N-1$.
This in turn implies that there is a fundamental problem with the work
by Gonzalez and Leal,\cite{GonzalezLeal} who have used $1/D$ expansion
to estimate energy levels of Lennard-Jones clusters in $D=3$ with
$N=3,4,\dots,7$ and $13$.  If such calculations could be done to
infinite order and re-summed to yield a convergent expression valid for
all $D$, the result would agree with this the analytic continuation
discussed above, but not with the physics of clusters with more than
four particles in three dimensions.

\appendix*
\section{}

Without loss of generality we can restrict our discussion to the
contribution to the transformed Hamiltonian $\H'$ of the kinetic
energy of particle $i=N$.  Correspondingly, we shall simplify
our notation as follows:
\begin{eqnarray}
&a_i=a_{N;i}\\
&g_{ij}=g_{N;ij}\\
&r_{i}=r_{Ni}
\end{eqnarray}
Note in particular that $r_i$ is {\em not} the distance of particle
$i$ to the origin, as suggested by convention and Eq.~(\ref{eq.r}),
but rather the distance of particle $N$ to particle $i$.

Define
\begin{equation}
\hat{\mat G}=(r_{i} g_{ij} r_{j} )_{i,j=1}^{N-1}.
\label{eq.ap1}
\end{equation}
Then the square of volume of the parallelepiped is given by the Grammian
\begin{equation}
\omega=\det({\hat{\mat G}}).
\label{eq.ap2}
\end{equation}

Consider a symmetric $s\times s$ matrix $\mat M$ of the form
\begin{equation}
m_{ij}=\left\{
\begin{array}{ll}
u_i(x_i)&\mbox{ if }i=j\\
v_{ij}(x_i,x_j)=v_{ji}(x_j,x_i)&\mbox{ if }i<j
\end{array}\right.
\label{eq.ap3}
\end{equation}
Since only row and column $i$ depend on $x_i$ this implies that
\begin{equation}
{\partial\, \det(\mat M)\over \partial x_i}=
\sum_{j=1}^N M_{ji} \left( {\partial u_i(x_i)\over \partial x_i}\delta_{ij}+
2{\partial v_{ij}(x_{i},x_{j})\over \partial x_i}(1-\delta_{ij})\right)
\label{eq.ap3.1}
\end{equation}
where $M_{ij}$ is the $(i,j)$ cofactor of $\mat M$.

Now
\begin{equation}
{\partial \hat{\mat G}\over \partial r_{p}}=r_{p} {\mat C}^{p},
\label{eq.ap4}
\end{equation}
$i=1,\dots,N-1$, where ${\mat C}^{p}$ has elements equal to unity on
the off-diagonal of row and column $p$ and equal for diagonal element
$p$, i.e.,
\begin{equation}
c_{ij}^{(p)}=\delta_{ip}+\delta_{jp}
\label{eq.ap5}
\end{equation}

\begin{equation}
{\partial \omega\over \partial r_{i}}=2 r_{i} \sum_{j=1}^{N-1}\hat{\mat 
G}_{ji}.
\label{eq.ap6}
\end{equation}
{From} this we find
\begin{equation}
\sum_i\hat{g}_{ji} {\partial \omega\over \partial r_{i}}{1\over  r_{i}}=2 
\omega
\label{eq.ap6.1}
\end{equation}
This can be verified by solving this last equation for ${\partial
  \omega\over \partial r_{i}}{1\over r_{i}}$ by means of Cramer's
rule, which indeed yields back Eq.~(\ref{eq.ap6}).  Thus we find
\begin{equation}
\sum_i g_{ji} {\partial \omega\over \partial r_{i}}=2 {\omega\over r_{j}}
\label{eq.ap7}
\end{equation}
from which it follows that
\begin{equation}
\chi=\omega^{(1-D)/4}
\label{eq.ap8}
\end{equation}
so that
\begin{equation}
\sum_{i=1}^{N-1} g_{ji} {\partial \chi\over \partial r_{i}}=
\half ({1-D}) {\chi\over r_{j}}=-\half \chi a_j
\label{eq.ap9}
\end{equation}
which shows that the linear differential operators vanish in $T_N$.

%To obtain from this the expression for $U_n$ or more generally for
%$U_i$ as given in Eq.~(\ref{eq.Und}), substitute the expression for
%$a_j$ obtained from Eq.~(\ref{eq.ap9}) in the first term on the right of
%Eq.~(\ref{eq.U}).  this yields a contribution to $U_n$ like the one in
%Eq.~(\ref{eq.Und}), but with coefficient $(d-1)^2/2$.

%By taking the derivative with respect to $r_{j}$ of Eq.~(\ref{eq.ap9}),
%and summing over $j$ we obtain an expression containing second-order
%derivatives of $\chi$, viz.
%\begin{equation}
%\sum_{i,j=1}^{n-1} g_{ij} {\partial^2 \chi\over \partial r_{i} \partial r_{j}}=
%-\half \sum_{j=1}^{n-1} {\partial ( \chi a_j )\over\partial r_{j}}-
%\sum_{i,j=1}^{n-1} {\partial g_{ij}\over\partial r_{i}}{\partial \chi\over \partial r_{j}}
%\end{equation}

To make the dependence on the spatial dimension $D$ explicit, we write
the first term on the right-hand side of Eq.~(\ref{eq.U}) as
\begin{equation}
\sum_{j=1}^{N-1} a_{j}\chi^{-1} {\partial\chi\over\partial r_j}=
-{1\over 8} (D-1)^2 \sum_{j=1}^{N-1} {1\over r_j}{\partial\log 
\omega\over\partial r_j}
\end{equation}

To calculate the second term we use the law of cosines in the form
\begin{equation}
g_{i;jk}={r_{ij}^2+r_{ik}^2-r_{jk}^2\over 2 r_{ij} r_{ik}},
\end{equation}
which yields
\begin{equation}
{\partial g_{i;jk}\over \partial{r_{ij}}}={1\over r_{ik}}-{1\over 
r_{ij}}g_{i;jk}
\end{equation}

By repeated use of Eq.~(\ref{eq.ap9}) we find
\begin{eqnarray*}
&&\sum_{j,k=1}^{N-1} g_{jk}\chi^{-1} {\ds\partial^2\chi\over\ds \partial 
r_{j} \partial r_{k}}=
\chi^{-1}\sum_{j,k=1}^{N-1} {\partial\over\partial 
r_j}\left(g_{jk}{\partial\chi\over\partial r_k}\right)
-\chi^{-1}\sum_{j,k=1}^{N-1} \left({1\over r_k}-{1\over r_j}g_{jk}\right) 
{\partial \chi\over\partial r_k}=\nonumber\\
&&-{1\over 8}(N-1)(D-1)\sum_{j=1}^{N-1} {1\over r_j}{\partial\log 
\omega\over\partial r_j}
\end{eqnarray*}

\begin{acknowledgments}
This research was supported by the United States National
Science Foundation under grant number ITR 0218858.
\end{acknowledgments}

%*********************************************************************************

%**********************************************************************************

\begin{thebibliography}{15}
\bibitem{footnote}Our
  treatment is related to the one given in
  Ref.~\onlinecite{AveryGoodsonHerschbach}. In comparing with their
  Section 3, it should be noted that our coordinate transformation
  from Cartesian coordinates to inter-particle distance has no
  inverse; nor do theirs, for that matter.
\end{thebibliography}
\end{document}